\documentclass[preprint,aps]{revtex4}
\usepackage{epsf}
\usepackage{graphicx} 
\usepackage{color} 

\begin{document}

\title
{
High-Order Coupled Cluster Method (CCM) Formalism 2 --  ``Generalised'' Expectation Values: 
Spin-Spin Correlation Functions for Frustrated and Unfrustrated 2D Antiferromagnets 
}

\author
{
D. J. J. Farnell
}
\affiliation
{Health Methodology Research Group, School of Community-Based Medicine, 
Jean McFarlane Building, University Place, University of Manchester,
Manchester M13 9PL, United Kingdom}

\date{\today}

\begin{abstract}
Recent developments of high-order CCM have been to extend existing formalism 
and codes to $s \ge \frac 12$ for both the ground and excited states. In this article,
we describe how ``generalised'' expectation values for a wide range of  
one- and two-body spin operators may also be determined using existing the CCM code 
for the ground state. We present new results for the spin-spin correlation functions
of the spin-half square- and triangular-lattice antiferromagnets by using the 
LSUB$m$ approximation. We show that the absolute values of the spin-spin 
correlation functions $| \langle \bf{s}(0).\bf{s}(r) \rangle |$ converge with increasing 
approximation level for both lattices. We believe that the LSUB$m$ approximation provides 
reasonable results for the correlation functions for lattice separations roughly of order 
$r \approx m$ for the square lattice. We compare qualitatively our results for the 
square lattice to those results of quantum Monte Carlo (QMC) and we see that good
correspondence is observed. Indeed as seen by QMC, the spin-spin correlation 
function initially decays strongly with $|r|$ before becoming constant for larger values of $|r|$. 
CCM results are also compared to results of exact diagonalisations for
both lattices. ED results demonstrate a strong finite-size effects at 
lattice separations $r=L/2$ (where $N=L \times L$) for both lattices. 
The behaviour of the correlation function for the triangular lattice is qualitatively 
similar to that of the square lattice, namely, that it decays strongly at first before
becoming constant. This is in keeping with the behaviour of both models, which 
are believed strongly to be N\'eel-ordered from approximate studies.
The CCM has been shown many times to provide consistently good results for 
the ground-state energy and sublattice magnetisation of a wide range of
quantum spin models. Here we have shown that the CCM also provides 
good results for the spin-spin correlation function.
\end{abstract}

\maketitle

\section{Introduction}

The coupled cluster method (CCM) \cite{refc1,refc2,refc3,refc4,refc5,refc6,refc7,refc8,refc9}  is a well-known method of quantum many-body theory (QMBT). The CCM has been applied with much success  in order to study quantum magnetic systems at zero temperature (see Refs. \cite{ccm1,ccm2,ccm999,ccm3,ccm4,ccm5,ccm6,ccm7,ccm8,ccm9,ccm10,ccm11,ccm12,ccm13,ccm13a,ccm14,ccm15,ccm16,ccm17,ccm18,ccm19,ccm19a,ccm20,ccm21,ccm22,ccm23,ccm24,ccm24a,ccm26,ccm27,ccm27a,ccm28,ccm29,ccm30,ccm31,ccm32,ccm33,ccm34,ccm35,ccm36,ccm37,ccm38,ccm39}). In particular, the use of computer-algebraic implementations \cite{ccm12,ccm15,ccm20,ccm39} of the CCM has been found to be very effective with respect to these spin-lattice problems. 
Recent developments of high-order CCM formalism and codes have been to treat systems 
with spin quantum number of $s \ge \frac 12$ for both the ground and excited states \cite{ccm39}.
In this article, we show how these ground-state formalism and codes may also be used directly 
to find ``generalised'' expectation values; that is, expectation values for a wide range of 
one- or two-body spin operator that are defined prior to the CCM calculation.

We apply the CCM to the spin-half Heisenberg model on the square and triangular lattices. The Hamiltonian is specified as follows,
\begin{equation}
H = \sum_{\langle i,j\rangle} {\bf s}_i ~ \cdot ~ {\bf s}_j
~~ ,
\label{heisenberg}
\end{equation}
where the sum on $\langle i,j\rangle$ counts all nearest-neighbour pairs once. The ground states of all of the cases considered here are classically ordered, albeit by a reduced amount due to quantum fluctuations. Indeed, the best estimates of the amount of classical ordering of the square lattice case 
from approximate methods are 61$\%$ from CCM\cite{ccm28}, 61.4$\%$ from quantum Monte Carlo studies\cite{qmc3}, 61.4$\%$ from series expansions\cite{series3}, 61.38$\%$ from spin-wave theory\cite{swt3}, and 63.4$\%$ from exact diagonalisations\cite{ed1}. Good results for the spin-spin
correlation function of the spin-half square-lattice Heisenberg antiferromagnet were found 
using quantum Monte Carlo in Ref. \cite{qmc4} on lattices of size up to $N=L \times L$, where $L=\{16,22,32\}$. They observed that the correlation functions decayed with separation $r$. 
However, they also saw finite-size effects,  indicated by cusp-like behaviour in these correlation functions, at distances given by $L/2$. Exact diagonalisations have also been carried out
for the correlation function of the square lattice \cite{ed2}. 
For the spin-half triangular-lattice antiferromagnet, the best estimates of the amount of 
classical ordering are 41$\%$ from quantum Monte Carlo studies\cite{qmc2}, 40$\%$ from 
series expansions\cite{series2}, 39$\%$ from exact diagonalisations\cite{ed1}, and 
40$\%$ from previous CCM calculations\cite{ccm28}. Very few results for the spin-spin 
correlation functions of the triangular lattice seem to exist apart from results of exact 
diagonalisations (e.g., see \cite{edt}).

\section{Method}

The details of the practical application of high-order coupled cluster method (CCM) formalism
to lattice quantum spin systems  are given in Refs. \cite{ccm12,ccm15,ccm20,ccm26,ccm39} and
also in the appendices to this article. However, we point out now that the 
ket and bra ground-state energy eigenvectors, $|\Psi\rangle$ and  $\langle\tilde{\Psi}|$, 
of a general many-body system described by a Hamiltonian $H$, are given by
\begin{equation} 
H |\Psi\rangle = E_g |\Psi\rangle
\;; 
\;\;\;  
\langle\tilde{\Psi}| H = E_g \langle\tilde{\Psi}| 
\;. 
\label{eq1} 
\end{equation} 
Furthermore, the ket and bra states are parametrized within the single-reference CCM as follows:   
\begin{eqnarray} 
|\Psi\rangle = {\rm e}^S |\Phi\rangle \; &;&  
\;\;\; S=\sum_{I \neq 0} {\cal S}_I C_I^{+}  \nonumber \; , \\ 
\langle\tilde{\Psi}| = \langle\Phi| \tilde{S} {\rm e}^{-S} \; &;& 
\;\;\; \tilde{S} =1 + \sum_{I \neq 0} \tilde{{\cal S}}_I C_I^{-} \; .  
\label{eq2} 
\end{eqnarray} 
One of the most important features of the CCM is that one uses a 
single model or reference state $|\Phi\rangle$ that is normalized.
We note that the parametrisation of the ground state has the normalization condition for the 
ground-state bra and ket wave functions ($\langle \tilde\Psi|\Psi\rangle 
\equiv \langle\Phi|\Phi\rangle=1$). The model state is required to have the 
property of being a cyclic vector with respect to two well-defined Abelian 
subalgebras of {\it multi-configurational} creation operators $\{C_I^{+}\}$ 
and their Hermitian-adjoint destruction counterparts $\{ C_I^{-} \equiv 
(C_I^{+})^\dagger \}$.  The interested reader is referred to the Appendices
and to Ref. \cite{ccm39} for more information regarding how the CCM problem 
is solved for.

Here, we use the classical ground state as the model state.  For the square lattice, 
this is the N\'eel state in which neighbouring spins are antiparallel, and, for the triangular lattice,
this a N\'eel-lke state in which neighbouring spins on three sublattices are at 
120$^\circ$ to each other.  For the square lattice, we perform a rotation of the local axes
of the up-pointing spins by 180$^\circ $ about the
$y$-axis. The transformation is described by,
\begin{equation}
s^x \; \rightarrow \; -s^x, \; s^y \; \rightarrow \;  s^y, \;
s^z \; \rightarrow \; -s^z  \; .
\end{equation}
The model state now appears $mathematically$ to consist
of purely down-pointing spins. 
In terms of the spin raising and lowering operators
$s_k^{\pm} \equiv s_k^x \pm {\rm i} s_k^y$ the Hamiltonian 
may be written in these local axes as,
\begin{equation}
H = -\frac 12 \sum_{\langle i,j \rangle}^N \; \biggl[ \; s_i^+
s_j^+ + s_i^-s_{j }^- + 2 s_i^z s_{j }^z  \; \biggr] \; ,
\label{eq:newH}
\end{equation}
where the sum on $\langle i,j\rangle$ again counts all nearest-neighbour pairs once
on the square lattice. 
Again, the classical ground-state of the Heisenberg model of Eq. (\ref{heisenberg}) 
for the triangular lattice is the N\'eel-like state where all spins on each 
sublattice are separately aligned (all 
in the $xz$-plane, say). The spins on sublattice A 
are oriented along the negative {\em z}-axis, and spins on sublattices 
B and C are oriented at $+120^\circ$ and $-120^\circ$, respectively, 
with respect to the spins on sublattice A.  We again rotate the local spin 
axes of those spins on the different sublattices. 
Specifically, we leave the spin axes on sublattice A unchanged, and we 
rotate about the $y$-axis the spin axes on sublattices B and C by 
$-120^\circ$ and $+120^\circ$ respectively, 
\begin{eqnarray} 
s_B^x \rightarrow -\frac{1}{2} s_B^x - \frac{\sqrt{3}}{2} s_B^z \;\;  &;& \;\; 
s_C^x \rightarrow -\frac{1}{2} s_C^x + \frac{\sqrt{3}}{2} s_C^z \;\; , 
\nonumber \\
s_B^y \rightarrow s_B^y \;\; &;& \;\; s_C^y \rightarrow s_C^y \;\; , 
\nonumber \\ 
s_B^z \rightarrow  \frac{\sqrt{3}}{2} s_B^x -\frac{1}{2} s_B^z \;\; &;& \;\; 
s_C^z \rightarrow -\frac{\sqrt{3}}{2} s_C^x -\frac{1}{2} s_C^z \;\; . 
\label{eq20}
\end{eqnarray} 
Once again, the model state now appears to consist
of purely down-pointing spins. 
We may rewrite Eq. (\ref{heisenberg}) in terms of spins defined in these 
local quantisation axes for the triangular lattice, such that 
\begin{eqnarray} 
H = \sum_{\langle i\rightarrow j\rangle}
\Bigl\{ 
&& 
-{1\over 2} s_i^z s_j^z
+\frac{\sqrt{3}}{4}
\left( s_i^z s_j^+ +s_i^z s_j^- -s_i^+ 
s_j^z- s_i^-s_j^z \right) \nonumber \\  
&& 
+\frac{1}{8}
\left( s_i^+s_j^- + s_i^- s_j^+ \right) 
-\frac{3}{8}
\left( s_i^+ s_j^+ + s_i^- s_j^- \right) 
\Bigl\}  
\;\; .
\label{eq21}
\end{eqnarray}
We note that the summation in Eq. (\ref{eq21}) again runs over nearest-neighbour
bonds, but now also with a {\it directionality} indicated by 
$\langle i \rightarrow j\rangle$, which goes from A to B, B to C, and 
C to A. 

The CCM formalism is exact in the limit of inclusion of
all possible multi-spin cluster correlations within 
$S$ and $\tilde S$, although this is usually impossible to achieve
practically.  
Hence, we generally make approximations in both $S$ and $\tilde S$.  The three most commonly employed approximation schemes previously utilised have been: (1) the SUB$n$ scheme, in which all correlations involving only $n$ or fewer spins are retained, but no further restriction is made concerning their spatial separation on the lattice; (2) the SUB$n$-$m$  sub-approximation, in which all SUB$n$ correlations spanning a range of no more than $m$ adjacent lattice sites are retained; and (3) the localised LSUB$m$ scheme, in which all multi-spin correlations over all distinct locales on the lattice defined by $m$ or fewer contiguous sites are retained. 
Another important feature 
of the method is that the bra and ket states are not always 
explicitly constrained to be Hermitian conjugates when we 
make such approximations, although the important 
Helmann-Feynman theorem is always preserved. We remark 
that the CCM provides results in the infinite-lattice limit 
$N \rightarrow \infty$ from the outset.

In this article, we wish to determine the spin-spin correlation functions
$| \langle \bf{s}(0).\bf{s}(r) \rangle |$  
as a function of lattice separation $|r|$ for the (unfrustrated) square-lattice
and (frustrated) triangular-lattice antiferromagnets.  We must take
into account the rotation of the local spin axes, although this proceeds
in exactly the same manner as for the Hamiltonian above. 
Again, we remark that the manner in which high-order CCM may be solved has
been discussed in Ref. \cite{ccm39}. The manner in which 
the ground-state CCM ket- and bra-state equations are solved 
is discussed in the Appendix to this article. In particular, the method by 
which ``generalised expectation values'' for a variety of one- and 
two-body spin operators may be obtained is explained in the
appendices.

\section{Results}

\begin{figure}
\epsfxsize=11cm
\centerline{\epsffile{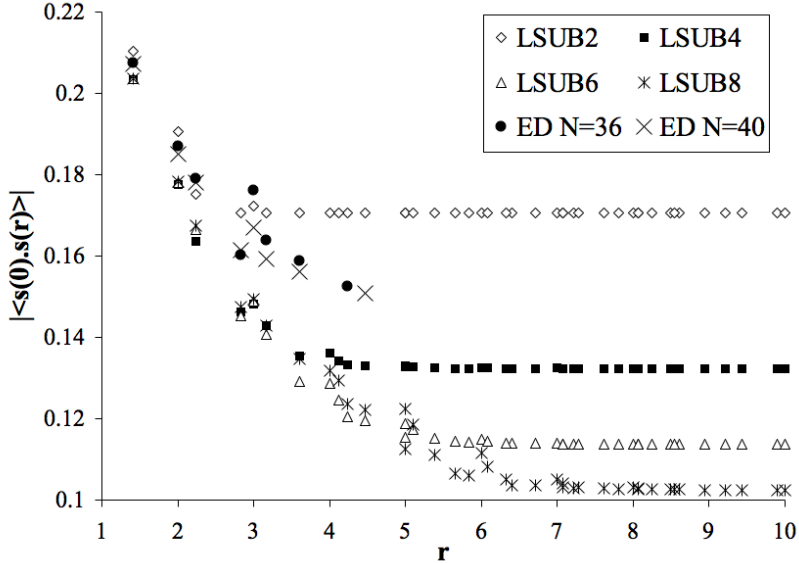}}
\caption{CCM results for the absolute values of the spin-spin correlation function for the spin-half 
square-lattice antiferromagnet  using the LSUB$m$ approximation with 
$m=\{2,4,6,8\}$. CCM results are compared to those results of exact diagonalisations (ED) 
with $N=36,40$ \cite{ed2}.}
\label{fig1}
\end{figure}

The results for the absolute values of the spin-spin correlation functions for the 
 (unfrustrated) spin-half square-lattice antiferromagnet using the LSUB$m$ 
approximation with $m = \{ 2,4,6,8 \}$ are shown in Fig. \ref{fig1}. The LSUB$m$ 
results are in  very good mutual agreement for small lattice separations for 
different values of $m$. Indeed, LSUB$m$ results are clearly converging with 
increasing levels of approximation.  Furthermore, we see that LSUB6 and LSUB8 
correspond reasonably well with each other up to $r  \stackrel {\sim}{<} 5$. From 
Fig. \ref{fig1} (as a ``rule of thumb'' only), we believe LSUB$m$ results ought to 
provide reasonable results to lattice separations approximately of order $m$ for
the square lattice.

CCM results are also compared to those results of exact diagonalisations (ED) \cite{ed2} 
in Fig. \ref{fig1}. However, we see that we obtain good correspondence only for very 
small lattice separations $r  <  3$. (We note that $N=40$ is at the current upper 
limit for ED of computational tractability.) However, the agreement between ED 
and CCM becomes rapidly worse for lattice separation $r \ge 3$. This disparity can be 
understood by considering results of quantum Monte Carlo (QMC) of Ref. \cite{qmc4}
for the spin-spin correlation function, which were carried out for for much larger lattices 
of size of $N=L \times L$, where $L = \{ 16,22,32  \}$. Indeed, we find good 
correspondence qualitatively (i.e., compared by by eye only) between the LSUB8 results of 
Fig. \ref{fig1} and those QMC results for the largest lattice $N = 32 \times 32$ of Fig. 3a in Ref. 
\cite{qmc4}. Furthermore, the authors of Ref. \cite{qmc4} noted a strong cusp-like behaviour 
was seen in the spin-spin correlation function at lattice separations of $L/2$ due to
finite-lattice effects. This is also seen in Fig. \ref{fig1}  for the ED results at $r=3$, although 
this cusp may be exacerbated by a small ``kink'' in the ``true'' spin-spin correlation  function
that occurs at this point anyway  (as seen in both CCM and QMC results). 
Interestingly,  we see from Fig. 3b of Ref. \cite{qmc4} that the accuracy of QMC results 
for the spin-spin correlation functions becomes more problematic (again as a
``rule of thumb'' only) for separations of order $r \stackrel {\sim}{>}  L/2$ for a 
given lattice size $N = L \times L$. Hence, we might also reasonably expect that 
results of ED might {\it only} be good for separations of $r < 3$ 
for lattices of size $N=36$ and $n=40$. Indeed, we note that ED agrees well with 
highly converged CCM results (and incidentally QMC results of  Fig. 3a in 
Ref. \cite{qmc4} -- again comparing by eye only) in this region. 

The author of the QMC calculations in Ref. \cite{qmc4} noted that: 
``to good approximation, the two-point  function depends only on 
$|r|$.'' As also noted in Ref. \cite{qmc4}, the long-range behaviour of 
correlation function gives a constant, thus indicating a long-range ordered 
ground state. We see this also at all levels of approximation as shown in Fig. \ref{fig1}.
Results \cite{ccm15} for the sublattice magnetisation have shown that extrapolation 
of the CCM LSUB$m$ results gives good correspondence to the results of QMC, as
noted above, where both methods indicate from the sublattice magnetisation
that approximately 61$\%$ of the classical N\'eel order remains in the quantum 
limit. Finally, the author of Ref. \cite{qmc4} remarked that the approach 
to the constant for larger separations could be fitted by either
an exponential or power-law form. We tried both for the LSUB8 
data with $r \le 8$ and found nothing to contradict this statement, although
the power-law decay seemed to work better (i.e., had a lower 
residual error) that the exponential law. The residual sum-of-squares
error for the power-law form was 0.0015 and for the exponential form was 0.0034. 
Fits of these forms to the LSUB8 data were carried out using the R statistics 
language: http://cran.r-project.org/.

\begin{figure}
\epsfxsize=11cm
\centerline{\epsffile{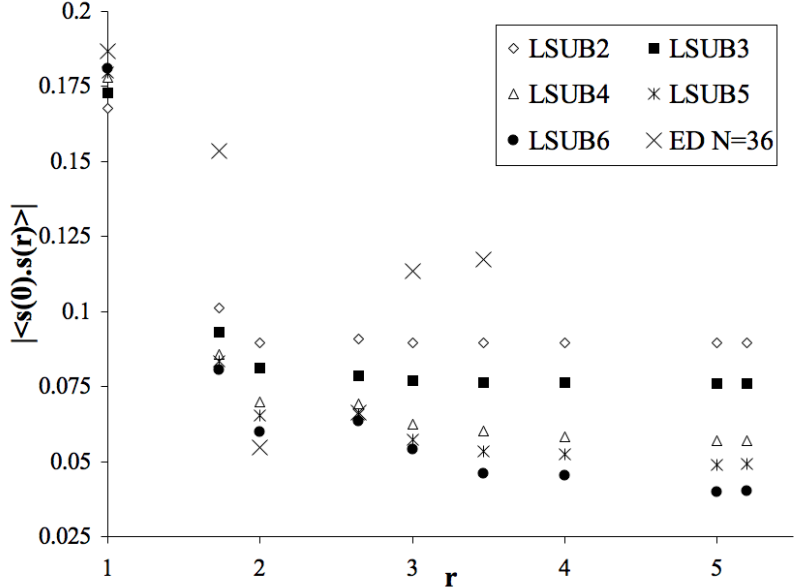}}
\caption{CCM results for the absolute values of the spin-spin correlation function for the spin-half 
triangular-lattice antiferromagnet  using the LSUB$m$ approximation with 
$m=\{2,3,4,5,6\}$. CCM results are compared to those results of exact diagonalisations 
(ED) \cite{edt}  with $N=36$.}
\label{fig2}
\end{figure}

The results for the (frustrated) spin-half square-lattice antiferromagnet using the LSUB$m$ 
approximation with $m = \{ 2,3,4,5,6 \}$ are shown in Fig. \ref{fig2}. Again, LSUB$m$
results are clearly converging with increasing levels of approximation, and again they are 
in very good mutual agreement for small lattice separations. CCM results are again 
compared to results of ED for$N=36$ (e.g., see Ref. \cite{edt}), and again good agreement 
is found for very small lattice separations ($r<3$). Another strong cusp-like behaviour is 
seen in ED results for the spin-spin correlation function for $r \approx 3$. Again, 
it is highly likely that this is primarily due to finite-size effects. Indeed, we see that 
CCM results for $r \ge 3$ are quite different to those of ED. Furthermore, CCM LSUB$m$
results initially decay strongly and then tend to a constant value for larger lattice separations. 
The behaviour of the correlation function for the triangular lattice is broadly similar 
 to that of the square lattice. Again, extrapolation of LSUB$m$ 
results \cite{ccm28} for the sublattice magnetisation demonstrate that the spin-half triangular 
lattice antiferromagnet is N\'eel ordered. For LSUB6 with $r \le 5.5$, we found that the
initial decay of the correlation function for CCM data in this case could be fitted well
by either an exponential or power-law form (for both forms, residual sum-of-squares: 0.0002).

\section{conclusions}

We have presented new formalism for the CCM in order to form 
``generalised'' expectation values for a wide range of terms in the spin
Hamiltonian. This is particularly useful for ``high-order'' CCM \cite{thecode}. 
We have applied this approach in order to determine the
spin-spin correlation functions for frustrated and unfrustrated 2D antiferromagnets,
namely, the spin-half triangular- and square-lattice Heisenberg models.
We found good correspondence of CCM with results of quantum Monte Carlo
(QMC) for the square-lattice antiferromagnet. A strong cusp-like behaviour was 
noted in exact diagonalisation (ED) results for the correlation function
at a lattice separation $L/2$ for a lattice of size $N = L \times L$ for both
lattices. This is a finite-size effect. A cusp in the spin-spin correlation function 
was also seen in QMC results at  $r=L/2$ for a $N = L \times L$ lattice. 

By contrast, this cusp was not seen in CCM results for either lattice at any value of $r$ 
at any level of LSUB$m$ approximation. Indeed, we note that CCM results are
determined in the infinite-lattice limit ($N \rightarrow \infty$) from the 
outset. We believe that (as a ``rule of thumb'') we obtain reasonable results 
for the correlation function up to separations of $r \approx m$ for the square lattice
-- even despite the fact that we are using a purely ``localised'' LSUB$m$ approximation 
scheme. LSUB$m$ results for the triangular lattice are clearly converging, and are
in good mutual agreement for small lattice separations. For both lattices, 
CCM results for the correlation function decay strongly initially, although 
they become constant for larger values of $r$. Previous CCM results for the sublattice
magnetisation \cite{ccm28} indicate that the spin-half Heisenberg models for the square
and triangular lattices are N\'eel ordered, which agrees with the results 
of other approximate methods.

It is still fair to say that QMC results still generally provide the 
most accurate results for the unfrustrated Heisenberg model on lattices of two 
spatial dimensions. However, QMC is severely restricted by the ``sign problem,'' 
which is a consequence of frustration at $T=0$. The CCM has been shown 
\cite{ccm28} to provides good results for the ground-state properties of
the spin-half triangular lattice antiferromagnet such as the ground-state
energy and sublattice magnetisation. Here we have shown that the 
CCM also provides reasonable results for the spin-spin correlation function
of the spin-half triangular lattice antiferromagnet. Finally, we note that
CCM results indicate that the behaviour of the correlation function for the triangular 
lattice is broadly similar to that of the square lattice.

\begin{acknowledgements}
We thank and acknowledge Joerg Schulenburg for the continuing development of the CCCM code;
excellent advice, help, and support over the years; and finally for providing exact diagonalisation 
results quoted in this manuscript \cite{ed2} using the SpinPack code (http://www.ovgu.de/jschulen/spin/)
\end{acknowledgements}

\pagebreak

\appendix

\section{CCM Formalism}
We begin the description of the application of the CCM by noting that it may be proven from
Eqs. (\ref{eq1}) and (\ref{eq2}) in a straightforward manner that the ket- and bra-state 
equations are thus given by
\begin{eqnarray} 
\langle\Phi|C_I^{-} {\rm e}^{-S} H {\rm e}^S|\Phi\rangle &=& 0 ,  \;\; 
\forall I \neq 0 \;\; ; \label{eq7} \\ 
\langle\Phi|\tilde{S} {\rm e}^{-S} [H,C_I^{+}] {\rm e}^S|\Phi\rangle 
&=& 0 , \;\; \forall I \neq 0 \;\; . \label{eq8}
\end{eqnarray}  
The index $I$ refers to a particular choice of cluster from the set of 
($N_F$) fundamental clusters that are distinct under the symmetries 
of the crystallographic lattice and the Hamiltonian and for a given 
approximation scheme at a given level of approximation.
We note that these equations are equivalent to the minimization 
of the expectation value of $\bar H = \langle \tilde \Psi | H | \Psi \rangle$ 
with respect to the CCM bra- and ket-state correlation coefficients 
$\{ \tilde{{\cal S}}_I, {\cal S}_I \}$. We note that Eq. (\ref{eq7}) is equivalent
to $\delta{\bar{H}} / \delta{\tilde{{\cal S}}_I}=0$, whereas Eq. (\ref{eq8}) is 
equivalent to $\delta{\bar{H}} / \delta{{\cal S}_I}=0$. Furthermore, we note 
that Eq. (\ref{eq7}) leads directly to simple form for the ground-state
energy given by
\begin{equation} 
E_g = E_g ( \{{\cal S}_I\} ) = \langle\Phi| {\rm e}^{-S} H {\rm e}^S|\Phi\rangle
\;\; . 
\label{eq9}
\end{equation}  
The full set $\{ {\cal S}_I, \tilde{{\cal S}}_I \}$ provides a complete 
description of the ground state. For instance, an arbitrary 
operator $A$ will have a ground-state expectation value given as 
\begin{equation} 
\bar{A}
\equiv \langle\tilde{\Psi}\vert A \vert\Psi\rangle
=\langle\Phi | \tilde{S} {\rm e}^{-S} A {\rm e}^S | \Phi\rangle
=\bar{A}\left( \{ {\cal S}_I,\tilde{{\cal S}}_I \} \right) 
\; .
\label{eq6}
\end{equation} 
The similarity transform of $A$ is given by,
\begin{equation}  
\tilde A \equiv {\rm e}^{-S} A {\rm e}^{S} = A 
+ [A,S] + {1\over2!} [[A,S],S] + \cdots 
\;\; .
\label{eq10}
\end{equation} 

\section{High-Order CCM}

The manner is which the ground-state problem is solved to high-orders
of approximation via computational is provided in Ref. \cite{ccm39}. However,
we note that use new ``high-order CCM operators'' that are formed purely
of spin-raising operators with respect to the model state. The model state
is taken to be a state in which all spins point in the downwards $z$-direction
after some appropriate rotation of the local axes of the spins. This allows 
us to write one- and two-body spin operators in terms of these new high-order
CCM operators. The problem inherent in Eq. (\ref{eq7}) becomes one of pattern-matching the fundamental 
set of clusters in $C_I^-$ to the terms in the similarity transformed 
version of the Hamiltonian $e^{-S} H e^S |\Phi\rangle$, where the $I$-th such equation is given by
\begin{equation}
E_I \equiv  \langle \Phi | C_I^- e^{-S} H e^{S} | \Phi \rangle = 0 ~~,
\forall I \ne 0 ~~ .
\label{tempLabel}
\end{equation}
(Note that we assume that $\langle \Phi | C_I^- C_I^+ | \Phi \rangle = 1$ in the above
equation). 
Specific terms in the Hamiltonian that may be used in the CCM code are: 
$s^z s^z$; $s^+ s^z$; $s^- s^z$;  $s^z s^+$; $s^z s^-$; $s^+ s^-$; $s^- s^+$;
$s^+ s^+$; $s^- s^-$; $s^+$; $s^-$; $s^z$; and, $(s^z)^2$.         
We now ``pattern-match''  the $C_i^-$ operators to those the relevant terms in the 
similarity transformed version of the Hamiltonian in order to form the CCM equations $E_I =0$ 
of Eq. (\ref{tempLabel}) at a given level of approximation. 

We now define the following {\it new} set of CCM bra-state correlation 
coefficients given by  $x_I\equiv {\cal S}_I$  and $\tilde{x}_I\equiv 
N_B/N (l!)\nu_I \tilde{\cal S}_I$  and we assume again that 
$ \langle \Phi | C_I^- C_I^+ | \Phi \rangle=1$. 
Note that $N_B$ is the number of Bravais lattice sites. Note also 
that for a given cluster $I$ then $\nu_I$ is a symmetry factor which 
is dependent purely on the point-group symmetries (and {\it not} 
the translational symmetries) of the crystallographic lattice and 
that $l$ is the number of spin operators. We note 
that the factors $\nu_I$, $N$, $N_B$, and $(l!)$ never need 
to be explicitly determined. 
The CCM bra-state operator may thus be rewritten as
\begin{equation}
\tilde S \equiv 1 + N \sum_{I=1}^{N_F} {\tilde x_I} C_I^- ~~ ,
\end{equation}
such that we have a particularly simple form for $\bar H$, given by
\begin{equation}
\bar{H} = N \sum_{I=0}^{N_F} \tilde x_I E_I ~~,
\label{appendix15}
\end{equation}
where $\tilde x_0=1$. 
We note that the $E_0$ is
defined by $E_0 = \frac 1N \langle \Phi| e^{-S} H e^S | \Phi \rangle$ 
(and, thus, $E_0=\frac 1N E_g$) 
and that $E_I$ is the $I$-th CCM ket-state equation defined
by Eq. (\ref{tempLabel}). The CCM ket-state equations are easily 
re-derived by taking the partial derivative of $\bar{H}/N$ with 
respect to $\tilde x_I$, where
\begin{equation}
\frac {\delta{(\bar{H}/N)}}{\delta \tilde x_I} (\equiv 0) = E_I ~~.
\label{appendix16}
\end{equation}
We now take the partial derivative of $\bar{H}/N$ with respect to 
$x_I$ such that the bra-state equations take on a particularly 
simple form, given by
\begin{equation}
\frac {\delta{(\bar{H}/N)}}{\delta x_I} =
\frac {\delta{E_0}}{\delta x_I} + 
\sum_{J=1}^{N_F} \tilde x_J \frac {\delta{E_J}}{\delta x_I} (\equiv 0) = \tilde E_I~~.
\label{appendix17}
\end{equation}

\section{Generalized Ground-State Expectation Values}

The expectation value of a ``generalized'' spin operator that we shall call $A$ 
may be treated in an analogous manner to that of the expectation value of the 
Hamiltonian, given by 
$\bar{H}$. We write:
\begin{equation}
A_I =  \langle \Phi | C_I^- e^{-S} A e^{S} | \Phi \rangle 
\label{tempLabel2}
\end{equation}
and with $C_0^-=1$. The similarity transform of $A$ is defined by Eq. (\ref{eq6}). 
We may treat a wide range of one- and two-body spin operations. However, unlike 
the Heisenberg Hamiltonian of Eq. (\ref{heisenberg}), we do not constrain $i$ 
and $j$ in the two-body terms to be only nearest-neighbors or 
next-nearest-neighbors. For example, we consider here the spin-spin correlation 
functions. The expectation value of the generalized
(spin) operator may again be written in a particularly simple form as:
\begin{eqnarray}
\bar A &=& \langle \tilde \Psi | A | \Psi  
\rangle  ~~  \nonumber \\
&=& N \sum_{I=0}^{N_F} {\tilde x_I} \langle \Phi 
| C_I^-  e^{-S} A e^{S} | \Phi \rangle  ~~  \nonumber \\
\Rightarrow \bar A&=& N \sum_{I=0}^{N_F} \tilde x_I A_I ~~ ,
\label{appendix18}
\end{eqnarray}
where $\tilde x_0=1$ also and $A_0=\frac 1N \langle \Phi | e^{-S} A e^S | \Phi \rangle$. 
The same code used to find ground-state 
equations may be used to find the generalized expectation values. 
Again we note that the index $I$ in Eq. (\ref{appendix18}) runs from zero to 
$N_F$. Again, we note that factors such as $N_B$ or $\nu_I$ etc. 
do not need to be determined explicitly because they cancel 
because of the definition of $\{\tilde x_I \}$ given above.
The techniques need to achieve a computational solution have been discussed
extensively elsewhere, and the interested reader is referred to Refs. \cite{ccm12,ccm15,ccm20,ccm26,ccm39}  for more information. However, clearly we see 
that the summation over all fundamental clusters involved in evaluating 
$\bar A$ is carried out readily once the ket- and bra-state equations of Eqs. 
(\ref{appendix16}) and (\ref{appendix17}) have been solved for.


\end{document}